\documentclass[a4paper,11pt]{article}
\usepackage{pos}
\usepackage{placeins}
\usepackage{graphicx}  
\usepackage{dcolumn}   
\usepackage{bm}        
\usepackage{eucal}
\usepackage{bigints}
\usepackage{graphics}
\usepackage[colorlinks=true]{hyperref}
\usepackage{physics}
\usepackage[bb=boondox]{mathalfa}
\usepackage{tikz}
\usetikzlibrary{quantikz2}
\usepackage{mathrsfs}
\usepackage{subcaption}

\DeclareMathOperator{\DiagCirc}{DiagCirc}

\title{Mixed State Variational Quantum Eigensolver for the Estimation of Expectation Values at Finite Temperature}

\author*[a,b]{Giuseppe Clemente}

\affiliation[a]{Deutsches Elektronen-Synchrotron DESY, Platanenallee 6, 15738 Zeuthen, Germany.}
\affiliation[b]{Dipartimento di Fisica dell'Universit\`a di Pisa and INFN --- Sezione di Pisa, Largo Pontecorvo 3, I-56127 Pisa, Italy.}

\emailAdd{giuseppe.clemente@unipi.it}

\abstract{We introduce a novel hybrid quantum-classical algorithm for the near-term computation of expectation values in quantum systems at finite temperatures.
This is based on two stages: on the first one, a mixed state approximating a fiducial truncated density matrix is prepared through Variational Quantum Eigensolving (VQE) techniques; 
this is then followed by a reweighting stage where the expectation values for observables 
of interest are computed. These two stages can then be iterated again with different hyperparameters to achieve arbitrary accuracy.
Resource and time scalability of the algorithm is discussed with a near-term perspective.}

\FullConference{The 40th International Symposium on Lattice Field Theory (Lattice 2023)\\
July 31st - August 4th, 2023\\
Fermi National Accelerator Laboratory\\}

\begin{document}
\maketitle

\section{Introduction}\label{sec:intro}
In the last few decades, standard Markov Chain Monte Carlo techniques 
proved successful in the investigation of non-perturbative features 
of lattice quantum field theories and non-abelian gauge theories such as QCD.
However, the classical simulation of real-time dynamical processes, 
phase diagram at non-zero baryonic chemical potential~\cite{Shapiro:1983,Rajagopal:2001,Philipsen:2010gj,Ding:2015,Aarts:2015tyj}, 
non-zero topological theta term~\cite{Shapiro:1983,Rajagopal:2001}, 
or frustrated spin systems~\cite{Wannier:1950zz,Balents,QuBiPF:2020iiz,Aiudi:2023cyq}, 
exhibit the infamous algorithmic sign problem.
In the future, there are serious hopes that these problems could be successfully tackled 
by new quantum algorithms tailored for quantum hardware and based on the Hamiltonian formulation.

Our interest is focused in particular on the estimation of \emph{thermal averages},
i.e., the expectation value of observables for a system at equilibrium at finite temperature.
Many quantum algorithms for thermal average estimation have been proposed 
in literature~\cite{Poulin_2009,Bilgin_2010,Riera_2012,Wu_2019,Zhu_2020,QMS_paper,QQMA_paper,Somma_Gibbs_Sampling,Moussa_2019,Motta_2020,Sun_2021,Lu_2021,Yamamoto:2022jes,Selisko_2022,Davoudi:2022uzo,Ball_2022,Powers_2023,Fromm_2023}.
As we discuss in more detail in Sec.~\ref{subsec:reweighting}, 
a single variational run of our algorithm can be used to extract information about any temperatures
by performing measurements of different observables at the reweighting stage.

This contribution is organized as follows: 
in Sec.~\ref{sec:algo} we give a general sketch of the algorithm,
which is based on two main stages. 
The variational stage for mixed-state preparation 
is described in Sec.~\ref{subsec:vqe}, 
while the reweighting stage for computing thermal averages is described in Sec.~\ref{subsec:reweighting}.
We show preliminary results of the application of this algorithm 
to a small transverse field Ising system in Sec.~\ref{sec:prenumres}, 
while in Sec.~\ref{sec:conclusions} we summarize and draw the conclusive remarks.

\section{Sketch of the Algorithm}\label{sec:algo}
Let us consider a system with Hamiltonian $H_0$ on a space $\mathcal{H}_{\text{sys}}$ with dimension $D$.
We introduce an auxiliary space $\mathcal{H}_{\text{aux}}$ of dimension $K\leq D$ such that the working 
space can be written as the tensor product between these two spaces
\mbox{$\mathcal{H}_{\text{tot}}=\mathcal{H}_{\text{aux}}\otimes \mathcal{H}_{\text{sys}}$}.
In terms of qubits, states in $\mathcal{H}_{\text{tot}}$ can be represented by 
two registers, one for the system and one for the auxiliary space, with a number of  
$q_S=\lceil \log_2 D\rceil$ and 
$q_A=\lceil \log_2 K \rceil$ qubits respectively,
while we denote the total number of qubits as $q_{\text{tot}}=q_S+q_A$.
The first variational stage of the algorithm, discussed in Sec.~\ref{subsec:vqe},
consists in preparing a pure state of the form 
\begin{align}\label{eq:partialmixstate}
    \ket{\Psi_{\vec{\gamma}}} = \sum_{k=0}^{K-1} \gamma_k \ket{k}\otimes \ket{\phi_k},
\end{align}
where the coefficients $\vec{\gamma}\equiv \{\gamma_k\}_{k=0}^{K-1}$ are constant parameters that can be freely chosen, with the only constraint that the resulting state is normalized (i.e., $\sum_{k=0}^{K-1} {|\gamma_k|}^2 = 1$),
while $\{\ket{\phi_k}\}_{k=0}^{K-1}$ indicate the first $K$ eigenvectors of $H_0$, associated to its non-decreasingly
ordered eigenvalues $\{E_k\}_{k=0}^{K-1}$. 
The state in Eq.~\eqref{eq:partialmixstate} represents a partially mixed state
in the form of a $K$-truncated density matrix when traced out on the auxiliary register as follows
\begin{align}\label{eq:truncatedrho_generic}
    \rho_{\vec{\gamma}}\equiv \Tr_{\mathcal{H}_{\text{aux}}} \Big[\ketbra{\Psi_{\vec{\gamma}}}{\Psi_{\vec{\gamma}}} \Big] =\sum_{k=0}^{K-1} {|\gamma_k|}^2 \ketbra{\phi_k}{\phi_k},
\end{align}
such that $w_k\equiv {|\gamma_k|}^2$ can be interpreted as the probability of projecting to the $k$-th eigenstate of $H_0$.
For their role in Eq.~\eqref{eq:truncatedrho_generic}, we refer to $\vec{\gamma}\equiv \{\gamma_k\}_{k=0}^{K-1}\in S^{K-1}_{\mathbb{C}}$ 
as the \emph{mixing coefficients}, 
while the auxiliary register will be referred as the \emph{mixing register} 
in the following discussions.
Notice that two choices of $\vec{\gamma}$ are equivalent if they represent the same density matrix, 
i.e., the complex phase of each coefficient is irrelevant.
Therefore, without loss of generality, we can assume to use real and positive mixing coefficients $\gamma_k>0 \;\forall k=0,\dots, K-1$.
Instead of aiming at the realization of Eq.~\eqref{eq:truncatedrho_generic} 
as an approximation to the density matrix of a system at finite temperature 
$\rho(\beta)\equiv \frac{e^{-\beta H_0}}{Z(\beta)}$  
(where $Z(\beta)=\Tr \big[e^{-\beta H_0}\big]$), we follow a different approach.
Indeed, for the estimation of expectation values at finite temperature,
one is interested in the evaluation of expectation values of hermitian observables $\mathcal{O}$,
realized as 
\begin{align}\label{eq:expvalO}
    \expval{\mathcal{O}}_{\mathcal{H}_{\text{sys}}}\!\!(\beta) &= \Tr_{\mathcal{H}_{\text{sys}}}\big[\mathcal{O}\rho(\beta)\big] = \lim_{K\to D} \frac{1}{Z^{(K)}(\beta)}\sum_{k=0}^{K-1} e^{-\beta E_k} \expval{\mathcal{O}}{\phi_k},
\end{align}
where $D=\dim \mathcal{H}_{\text{sys}}$ is the dimension of the physical space
and $Z^{(K)}(\beta) = \sum_{k=0}^{K-1} e^{-\beta E_k}$.
In terms of the arbitrary mixed state in Eq.~\eqref{eq:partialmixstate}, we can represent 
an approximation to the expectation value of Eq.~\eqref{eq:expvalO} 
introducing a linear functional operator $\mathcal{F}^{(\beta,\vec{\gamma})}[\mathcal{O}]$ such that
\begin{equation}\label{eq:intermediateExtension}
\begin{aligned}
    \expval{\mathcal{O}}_{\mathcal{H}_{\text{sys}}}\!\!(\beta)
    =\lim_{K\to D} \frac{\mel{\Psi_{\vec{\gamma}}}{\mathcal{F}^{(\beta,\vec{\gamma})}[\mathcal{O}]}{\Psi_{\vec{\gamma}}}}{\mel{\Psi_{\vec{\gamma}}}{\mathcal{F}^{(\beta,\vec{\gamma})}[\mathbb{1}]}{\Psi_{\vec{\gamma}}}}
    = \lim_{K\to D} \frac{\sum_{k,p=0}^{K-1} \gamma_k \gamma_p^* \Tr_{\mathcal{H}_{\text{tot}}} \Big[\mathcal{F}^{(\beta,\vec{\gamma})}[\mathcal{O}] \ketbra{k}{p} \otimes \ketbra{\phi_k}{\phi_p}\Big]}{\sum_{k,p=0}^{K-1} \gamma_k \gamma_p^* \Tr_{\mathcal{H}_{\text{tot}}} \Big[\mathcal{F}^{(\beta,\vec{\gamma})}[\mathbb{1}] \ketbra{k}{p} \otimes \ketbra{\phi_k}{\phi_p}\Big]},
\end{aligned}
\end{equation}
where the dependence on $\beta$ of $\mathcal{F}^{(\beta,\vec{\gamma})}$ and possibly also $\ket{\Psi^{(K)}}$ 
is implicitly understood.
In order to match the expressions in Eq.~\eqref{eq:expvalO} and Eq.~\eqref{eq:intermediateExtension}, one can impose 
\begin{align}\label{eq:reweighting_condition}
    \mathcal{N} e^{-\beta E_k}\delta_{k,p} \mel{\phi_k}{\mathcal{O}}{\phi_p}\overset{!}{=} \gamma_k \gamma_p^* \mel{k,\phi_k}{\mathcal{F}^{(\beta,\vec{\gamma})}[\mathcal{O}]}{p,\phi_p},
\end{align}
for some scalar constant $\mathcal{N}\neq 0$ (this factor is irrelevant and cancels out when we take the ratio with the expectation value with $\mathcal{F}^{(\beta,\vec{\gamma})}[\mathbb{1}]$).
This is possible with a functional operator of the form 
\begin{align}\label{eq:bigfunctional}
\mathcal{F}^{(\beta,\vec{\gamma})}[\mathcal{O}]\equiv \mathcal{N} A^{(\beta,\vec{\gamma})}\otimes \mathcal{O},
\end{align}
where the operator $A^{(\beta,\vec{\gamma})}$, which we refer to as the \emph{reweighting operator},
must satisfy
\begin{align}\label{eq:Ac_def}
   \mathcal{N} \gamma_k \gamma_p^* A^{(\beta,\vec{\gamma})}_{k,p} = e^{-\beta E_k}\delta_{k,p} 
   \implies A^{(\beta,\vec{\gamma})}_{k,p} \propto \frac{e^{-\beta E_k}}{{|\gamma_k|}^2}\delta_{k,p}.
\end{align}
At this point, we can use the freedom available in the state preparation stage, subsumed into the mixing coefficients $\vec{\gamma}$. 
If exact prior information on the spectrum is known, 
one can completely encode the thermal information directly into the mixing coefficients $\vec{\gamma}$, 
while the reweighting operator in Eq.~\eqref{eq:Ac_def} becomes trivially multiple of the identity.
Different prior choices of $\vec{\gamma}$ give in principle the same results for the thermal averages 
according to Eq.~\eqref{eq:intermediateExtension}, 
but there is a tradeoff in the numerical effort between 
the efficiency of convergence for the VQE stage used in the mixed state preparation 
and the accuracy of the reweighting procedure. 
It is also possible to build a dynamical sequence of mixing coefficients which improve the efficiency
for one or the other stage of the computation as needed for convenience (e.g., by changing adiabatically some parameters).

In the next Sections, we discuss how to implement in practice the state preparation via variational
methods (Sec.~\ref{subsec:vqe}) and a generic Hermitian operator thermal average via reweighting (Sec.~\ref{subsec:reweighting}).

\subsection{Mixed state preparation via VQE}\label{subsec:vqe}
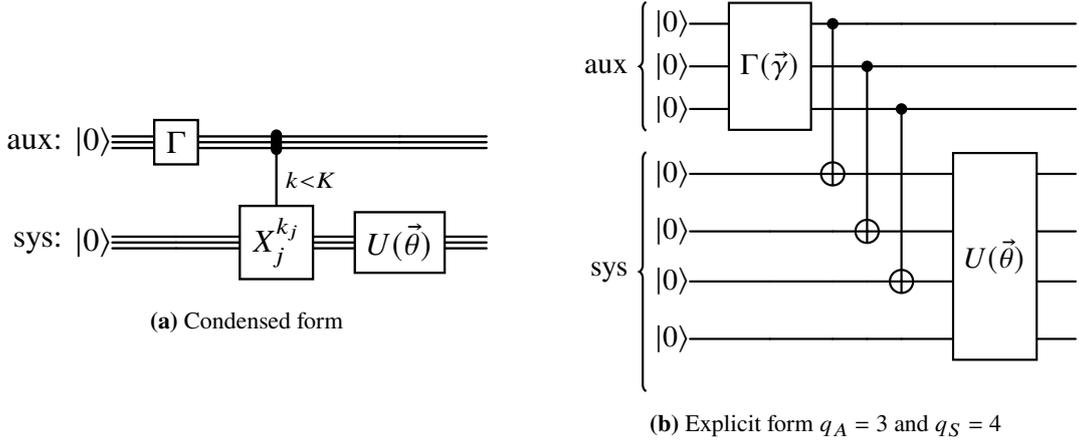
\begin{figure}[h]
\begin{subfigure}[t]{0.5\textwidth}
\centering
\resizebox{.9\textwidth}{!}{
\begin{quantikz}[wire types={b,b},classical gap=0.07cm]
 \lstick{\text{aux:} } \ket{0}& \gate{\Gamma} & \ctrl[wire style={"k<K"},vertical wire=q]{1} & &\\
 \lstick{\text{sys:} } \ket{0}& & \gate{X_j^{k_j}} & \gate{U(\vec{\theta})} &
\end{quantikz}
}
\caption{Condensed form}
\label{fig:unifAnsatzState_subfigA}
\end{subfigure}
\begin{subfigure}[t]{0.5\textwidth}
\centering
\resizebox{.9\textwidth}{!}{
\begin{quantikz}[row sep=0.3cm]
 \lstick[3]{\text{aux}} {\ket{0}}& \gate[3,disable auto height]{\Gamma(\vec{\gamma})} & [-1em] \ctrl{3} &[-1em] &[-1em] & &\\[-0.5em]
                        {\ket{0}}&  &&\ctrl{3} && &\\[-0.5em]
                        {\ket{0}}&  &&&\ctrl{3}& &\\
 \lstick[5]{\text{sys}} {\ket{0}}&  &\targ{}&&& \gate[4, disable auto height]{U(\vec{\theta})}&\\
                        {\ket{0}}&  &&\targ{}&& &\\
                        {\ket{0}}&  &&&\targ{}& &\\
                        {\ket{0}}&  && && &\\
\end{quantikz}
}
\caption{Explicit form $q_A=3$ and $q_S=4$}
\label{fig:unifAnsatzState_subfigB}
\end{subfigure}
\caption{Example of a mixed state ansatz preparation with mixing $\vec{\gamma}$.}
    \label{fig:unifAnsatzState}
\end{figure}

Let us consider a truncation with $q_A$ auxiliary qubits and $K=2^{q_A}$ levels.
Starting from the standard initialization $\ket{0}=\ket{0}_{\text{aux}}\otimes \ket{0}_{\text{sys}}$, 
we can first prepare the mixing register with a circuit $\Gamma$ such that $\Gamma \ket{0}_{\text{aux}} = \sum_{k=0}^{K-1} \gamma_k \ket{k}_{\text{aux}}$.
The procedure that we adopt to prepare the full state encodes the eigenstates associated 
to the lowest eigenvalues of $H_0$ by means of a Variational Quantum Eigensolver (VQE) approach~\cite{Peruzzo:2013bzg}. 
Broadly speaking, in its standard formulation, the VQE algorithm consists in parameterizing the state with a circuit $|\Psi(\vec{\theta})\rangle=U(\vec{\theta})\ket{0}$ and minimizing a cost functional, 
described as the expectation value of a Hermitian operator $\mathbb{H}$ as 
\begin{align}\label{eq:naiveCost}
    C\big[|\Psi(\vec{\theta})\rangle\big] = \langle\Psi(\vec{\theta}) | \mathbb{H} | \Psi(\vec{\theta}) \rangle,
\end{align}
where the minimization is done with respect to the parameters $\theta$.
The optimal state $|\Psi(\vec{\theta}^*)\rangle$ would then approximate the ground state of the 
operator $\mathbb{H}$, while $C(\vec{\theta}^*)$ approximates its corresponding eigenvalue.
Regarding our case, we can parameterize the state in such a way as not to spoil the mixing coefficients\footnote{besides an irrelevant complex phase coming from kickback effects.} 
$\gamma_k$ through an oracle ansatz that brings it into the form
\begin{align}\label{eq:fullParamPsiOrthoHard}
    |\Psi_{\vec{\gamma}}(\vec{\theta})\rangle = \sum_{k=0}^{K-1} \gamma_k \ket{k}_{\text{aux}}\otimes (U_k(\vec{\theta})\ket{0}_{\text{sys}}).
\end{align}
The Hermitian operator $\mathbb{H}$ acts on $\mathcal{H}_{\text{tot}}$ and one should relate it
to the original Hamiltonian $H_0$ acting on $\mathcal{H}_{\text{sys}}$ in order
to prepare the state in Eq.~\eqref{eq:partialmixstate} with the eigenstates of $H_0$ as target
states in the system register.
There is freedom in choosing this Hermitian operator but, for simplicity, 
in the following, we consider the simplest choice, given by $\mathbb{H}=\mathbb{1}\otimes H_0$.
In principle, each ansatz sub-circuit $U_k$ can be implemented by different parameterized circuits.
However, using completely independent circuits with an Hermitian operator such as
$\mathbb{H}=\mathbb{1}\otimes H_0$ is not sufficient to prepare 
the target state expressed by Eq.~\eqref{eq:partialmixstate},
since the cost would be minimized 
by a direct product state depending only on the ground state in the form
$\ket{\Psi} = \ket{\text{any}}\otimes \ket{\phi_0}$. 
Indeed, in this unconstrained situation, each $U_k$ would attain the state 
of minimum energy $U_k(\vec{\theta}^*)\ket{0}_{\text{sys}}\simeq \ket{\phi_0}$.
For this reason, it is useful to hard-code the orthogonality between
probing states $|\psi_k(\vec{\theta})\rangle \equiv U_k(\vec{\theta})\ket{0}$
by using a controlled initialization of the form $U_k(\vec{\theta})=U(\vec{\theta}) I_{k}$,
with $I_k \equiv {(\sigma^x)}^{\otimes {(k)}_2}= \bigotimes_{i=0}^{q_S} {(\sigma^x)}^{k_i}$,
where $k_i$ is the $i$-th binary digit of the integer $k$;
the action of $I_k$ on the standard initialized state
is essentially equivalent to writing the associated integer
number $k$ on the system register, since
$I_k \ket{0}_{\text{sys}}=\ket{k}_{\text{sys}}$.
An example of this initialization
is shown in Fig.~\ref{fig:unifAnsatzState}.
The orthogonality between states with different integer $k$ in Eq.~\eqref{eq:fullParamPsiOrthoHard} is 
then guaranteed by the initialization, since \mbox{$\langle \psi_k(\vec{\theta})|\psi_p(\vec{\theta})\rangle=\mel{k}{U^\dagger(\vec{\theta})U(\vec{\theta})}{p}=\delta_{k,p}$}.
The ansatz built as in Eq.~\eqref{eq:fullParamPsiOrthoHard}
will then minimize the cost function in Eq.~\eqref{eq:naiveCost},
obtaining a set of mutually orthogonal states $\{\ket{\psi_k}\}_{k=0}^{K-1}$ with least average energy,
i.e., ideally the first $K$ eigenstates of the Hamiltonian
$\{\ket{\phi_k}\}_{k=0}^{K-1}$.
The structure of this ansatz circuit is similar to the one recently described 
in~\cite{Consiglio:2023mev,Consiglio:2023aql} but, in our case, we keep 
the mixing completely arbitrary and perform the actual Gibbs preparation through reweighting, 
as described in Sec.~\ref{subsec:reweighting}.
The cost functional evaluated on the state in Eq.~\eqref{eq:fullParamPsiOrthoHard} 
then takes the form
\begin{align}\label{eq:explicitCost}
    C\big[\vec{\theta}; \vec{\gamma}\big] 
    = \sum_{k=0}^{K-1} {|\gamma_k|}^2 \mel{k}{U^\dagger(\vec{\theta})H_0 U(\vec{\theta})}{k}_{\text{sys}}.
  \end{align}
If we assume mixing coefficients $\{\gamma_k\}$ decreasing in modulo, 
the ideal minimum of the functional in Eq.~\eqref{eq:explicitCost} is realized by a unitary matrix $U(\vec{\theta})$ containing
the first $K$ lowest eigenvectors as columns ${[U(\vec{\theta})]}_{i,k} = \braket{i}{\phi_k} \; \forall k=0,\dots,K-1 \;\forall i=0,\dots,D-1$, which saturates the cost at the value
$C_{\text{min}}[\vec{\gamma}] = \underset{\vec{\theta}}{\min}\; C[\vec{\theta};\vec{\gamma}] = \sum_{k=0}^{K-1} {|\gamma_k|}^2 E_k$.

\subsection{Reweighting}\label{subsec:reweighting}
Let us assume we have proceeded through the first variational stage so that 
we prepared a state according to Eq.~\ref{eq:partialmixstate},
for some fixed mixing coefficients $\vec{\gamma}$. 
The next step is then to evaluate expectation values
of the functional in Eq.~\eqref{eq:bigfunctional} encoding a reweighting of the observable 
$\mathcal{O}$, as expressed by Eq.~\eqref{eq:Ac_def}.
As described in Sec.~\ref{sec:algo}, the thermal expectation value
of an operator $\mathcal{O}$ on $\mathcal{H}_{\text{sys}}$ can be estimated as 
the ratio between the expectation values of the functional operators 
$\mathcal{F}^{(\beta,\vec{\gamma})}[\mathcal{O}]$ and  $\mathcal{F}^{(\beta,\vec{\gamma})}[\mathbb{1}]$ on the extended space 
$\mathcal{H}_{\text{tot}}$.
The operator $\mathcal{F}^{(\beta,\vec{\gamma})}[\mathcal{O}]$ is already diagonal on the mixing register, 
but it involves the evaluation of the energies $E_k$, while a measurement on 
the observable side must be decomposed into a sum of easily diagonalizable pieces 
$\mathcal{O} = \sum_{m=0}^{M^{(\mathcal{O})}-1} S_m^{(\mathcal{O})\dagger} \Lambda^{(\mathcal{O})}_m S^{(\mathcal{O})}_m$,
where $\DiagCirc(\mathcal{O})\equiv \{(\Lambda^{(\mathcal{O})}_m,S^{(\mathcal{O})}_m)\}_{m=0}^{M^{(\mathcal{O})}-1}$ corresponds to pairs of diagonal 
Hermitian operators and diagonalizing unitaries 
(efficiently implemented as circuits) respectively, as customary in VQE applications.
The functional is then computed using at least $M_{\mathcal{O}}$ independent circuit evaluations: 
\begin{align}\label{eq:FO_exactEks}
\mathcal{F}^{(\beta,\vec{\gamma})}[\mathcal{O}]=\sum_{m=0}^{M^{(\mathcal{O})}-1} {(\mathbb{1}\otimes S^{(\mathcal{O})}_m)}^\dagger \Bigg(\sum_{k=0}^{K-1}\frac{e^{-\beta E_k}}{{|\gamma_k|}^2}\ketbra{k}{k}\otimes \Lambda^{(\mathcal{O})}_m\Bigg)(\mathbb{1}\otimes S^{(\mathcal{O})}_m)
\end{align}
However, the values of the eigenstates $E_k$ are not directly accessible in this form.

From the state after VQE, we can infer information about the lowest part of the spectrum
by measuring with a vector-valued Hermitian operator 
on $\mathcal{H}_{\text{tot}}$ defined as 
\begin{align}\label{eq:spectrumvec}
    \vec{\mathcal{E}} 
    \equiv \sum_{k=0}^{K-1} \frac{1}{{|\gamma_k|}^2}\hat{e}_k \ketbra{k}{k}_{\text{aux}} \otimes H_0 
=\sum_{i=0}^{M^{(H_0)}-1} {(\mathbb{1}\otimes S^{(H_0)}_i)}^\dagger \Bigg[\sum_{k=0}^{K-1} \frac{1}{{|\gamma_k|}^2}\hat{e}_k \ketbra{k}{k}_{\text{aux}} \otimes \Lambda^{(H_0)}_i\Bigg](\mathbb{1}\otimes S^{(H_0)}_i),
\end{align}
where $\hat{e}_k$ is the $k$-th unit vector in the standard basis
and we used a diagonal-circuit decomposition of the Hamiltonian $\DiagCirc(H_0) = \{(\Lambda^{(H_0)}_i,\hat{S}^{(H_0)}_i)\}_{i=0}^{M^{(H_0)}-1}$.
We call $\vec{\mathcal{E}}$ the \emph{spectrum vector}.
It is straightforward to show that, on the partially mixed state $\vec{\gamma}$ 
(with decreasing coefficients), its expectation value
reduces to the vector containing the first $K$ lowest eigenvalues as components, 
namely
\begin{align}\label{eq:spectrumvec_expval}
    \vec{\mathcal{E}}\equiv \mel{\Psi_{\vec{\gamma}}}{\vec{\mathcal{E}}}{\Psi_{\vec{\gamma}}}=\sum_{k=0}^{K-1} 
    E_k \hat{e}_k = {(E_0, E_1, \dots, E_{K-1})}^T\in \mathbb{R}^{K}.
\end{align}
Then, $E_k$ are estimated from the circuit and plugged into Eq.~\eqref{eq:FO_exactEks},
which is then used in Eq.~\eqref{eq:intermediateExtension}
to estimate the thermal average for any observable (energy included) at any value of $\beta$.

\section{Preliminary numerical results}\label{sec:prenumres}
\begin{figure}[h!]
  \centering
  \includegraphics[width=0.99\textwidth]{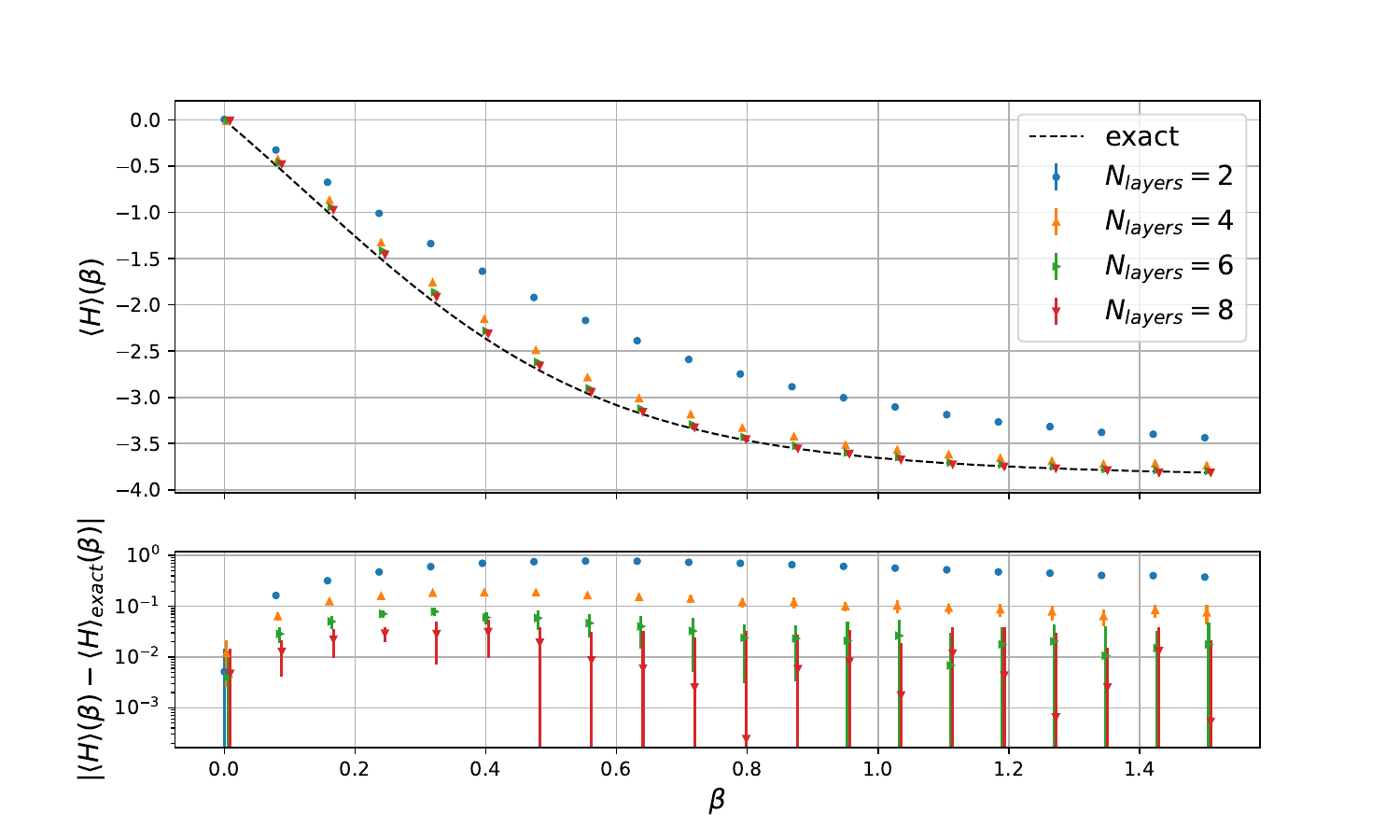}
  \caption{Thermal average of the energy for the transverse field ising model discussed in the text.
  The results are obtained through the reweighting stage at different values of $\beta$, 
  using the same state prepared via the VQE stage using a generic ansatz 
and for different numbers of layers. Points with different numbers of layers have been slightly shifted
on the x-axis for better readability.}
  \label{fig:tfising}
\end{figure}

\begin{figure}[h!]
  \centering
  \includegraphics[width=0.97\textwidth]{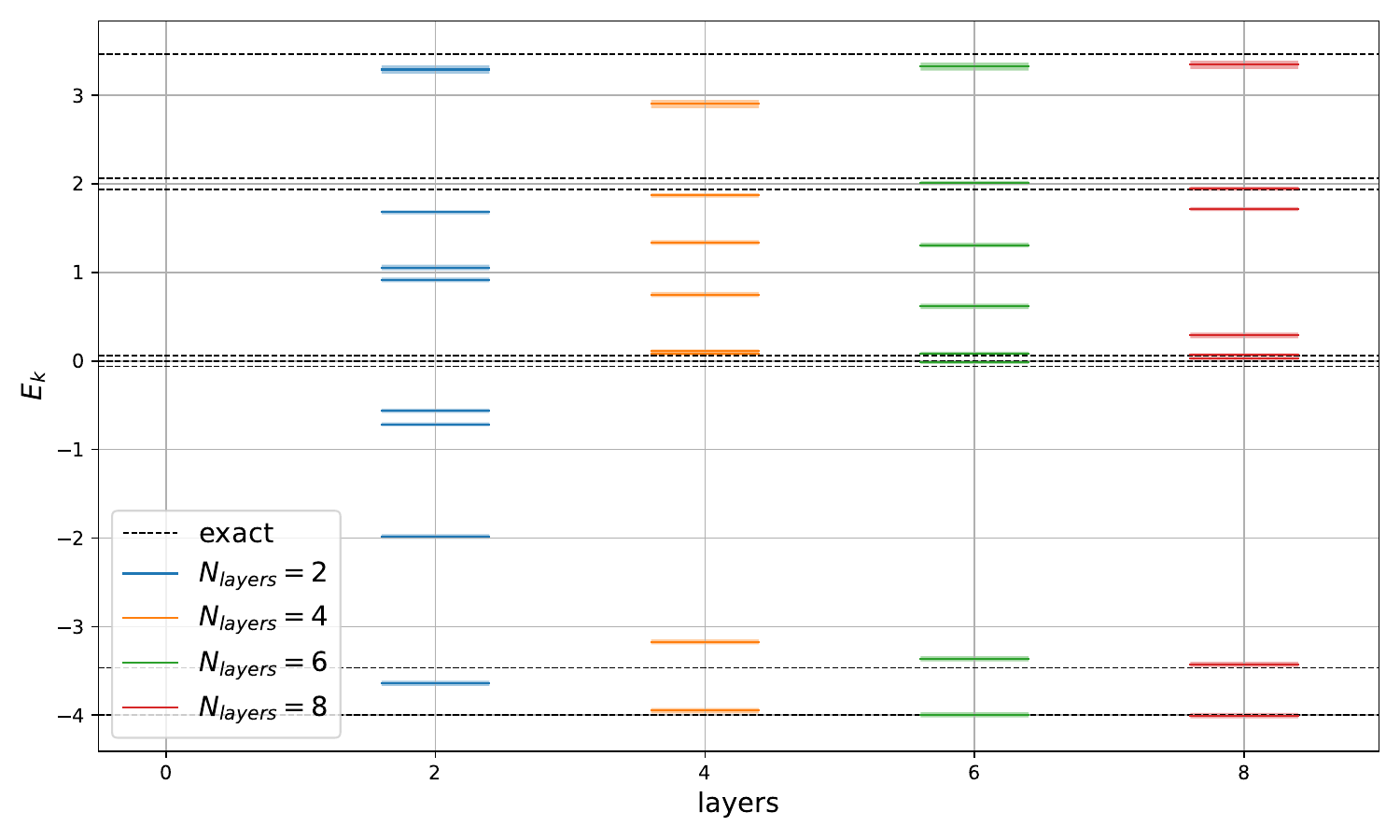}
  \caption{Estimated spectrum for the same runs shown in Fig.~\ref{fig:tfising}.
  Degenerate levels in the exact spectrum (dashed line) have been slightly split for clarity.}
  \label{fig:tfising_spectr}
\end{figure}

Some preliminary results of the application of the algorithm described above
are shown in Fig.~\ref{fig:tfising} for a one-dimensional transverse field Ising model
with Hamiltonian
\begin{align}\label{eq:H0_tfi}
    H=-J \sum_{i=0}^{q_S-1} \sigma_i^X \sigma_{(i+1)\!\!\!\!\!\mod q_S}^X - h \sum_{i=0}^{q_S-1} \sigma_i^Z,
\end{align}
where we considered in particular the case $q_S=3$, $q_A=3$, $J=1$ and $h=1$.
The ansatz $U(\vec{\theta})$ we chose for this proof of principle is generic,
made in particular of single layers of $X$ and $Z$ single-qubit parametric rotations, 
alternated with entangling layers made of CNOTs. 
While a tailored ansatz enforcing specific symmetries can be considered, 
we recall that the sub-circuits $\{U(\vec{\theta})\ket{k}_{\text{sys}}\}_{k=0}^{K-1}$ represent 
an orthogonal $K$-frame in the full Hilbert space, 
ultimately converging to the first $K$ lowest eigenstates 
of the Hamiltonian $H_0$, shown in Fig.~\ref{fig:tfising_spectr}
for different number of layers.

In terms of the mixing coefficients $\vec{\gamma}$,
the final results are in general robust under
different choices, but we observed variable performances during
the VQE stage in reaching the target state preparation expressed in Eq.~\eqref{eq:partialmixstate}.
In particular, for a mixing $\vec{\gamma}$ too close to the uniform one,
the cost function in Eq.~\eqref{eq:explicitCost} 
is only weakly dependent on the parameters $\vec{\theta}$, making the cost gradient too small and
the circuit untrainable towards the target state.
On the other hand, a mixing with too steep decreasing coefficients ${|\gamma_k|}^2\gg {|\gamma_{k+1}|}^2$
gives a good resolution for the energy associated to the ground state $|\psi_0(\vec{\theta})\rangle$,
virtually equivalent to the application of the standard VQE algorithm for the ground state only, 
but it yields less accurate results for all the other excited states, 
making their contributions to the cost in Eq.~\eqref{eq:explicitCost} undersampled and noisier, which affect also the final results of the reweighting stage in Eq.~\eqref{eq:intermediateExtension}.
A compromise between these to extreme cases realizes a variable accuracy as a function
of $\beta$ after reweighting, but the freedom in the mixing choice 
can be used to optimize the whole procedure adaptively, as mentioned in Sec.~\ref{sec:algo}.

\section{Conclusions}\label{sec:conclusions}
We introduced a novel algorithm for the evaluation of thermal 
averages through a generic mixed state preparation joined to a reweighting stage.
This combination allows for great flexibility in moving resources and computational effort 
between the two stages of the algorithm, 
depending on the desired target in both accuracy and efficiency.
A thorough discussion of the systematical and statistical error analysis, 
resource estimates, algorithmic improvements and variations will be presented in a future work.
In particular, we plan to investigate the behavior of the algorithm for larger-scale systems 
and for toy models of lattice gauge theories
such as the one considered in Ref.~\cite{Ballini:2023ljs,Ballini:2024wbe}. 
Also, we would investigate how the performance of the algorithm is affected by quantum noise on near-term hardware, and which types of error mitigation techniques~\cite{Charles:2023zbl,Zambello:2024rzj} are effective.

\section*{Acknowledgments}
I am grateful to Claudio Bonati, Massimo D'Elia, 
Lorenzo Maio and Kevin Zambello for their valuable feedback 
and for being sources of encouragement during the writing of this manuscript. 

\bibliographystyle{JHEP}
\bibliography{references.bib}

\end{document}